\documentstyle[12pt]{article}
\newcommand{\fnd}[2]{\frac{\textstyle #1}{\textstyle #2}}
\newcommand{\xrm}[1]{{\textstyle \mbox{\rm #1}}}
\newcommand{\ket}[1]{\mbox{$\left| #1\right\rangle$}}
\newcommand{\bk}[2]{\mbox{$\left\langle #1\left| #2\right.\right\rangle$}}
\newcommand{\bVk}[3]{\mbox{$\left\langle #1\left| #2\right| #3\right\rangle$}}
\begin{document} \baselineskip .7cm
\title{Flavour-symmetry correction to the ``na\"{\i}ve'' Zweig rule for the
scalar-meson flavour-singlet}
\author{
Eef van Beveren\\
{\normalsize\it Departamento de F\'{\i}sica, Universidade de Coimbra}\\
{\normalsize\it P-3000 Coimbra, Portugal}\\
{\small eef@malaposta.fis.uc.pt}\\ [.3cm]
\and
George Rupp\\
{\normalsize\it Centro de F\'{\i}sica das Interac\c{c}\~{o}es Fundamentais}\\
{\normalsize\it Instituto Superior T\'{e}cnico, Edif\'{\i}cio Ci\^{e}ncia}\\
{\normalsize\it P-1096 Lisboa Codex, Portugal}\\
{\small george@ajax.ist.utl.pt}\\ [.3cm]
{\small PACS number(s): 14.40.Cs, 12.39.Pn, 13.75.Lb}\\ [.3cm]
{\small hep-ph/9902301}
}
\date{\today}
\maketitle
\begin{abstract}
We present flavour-symmetric couplings for the OZI-allowed three-meson
vertices of effective meson theories, which for the case of the two-meson
channels to which the flavour-singlet scalar meson couples, are endowed with
a correction factor with respect to the standard formula.
\end{abstract}
\clearpage

\section{Introduction}

Three-meson vertices give rise to the most important strong interactions which
are considered by effective meson theories, as they reflect the simple fact
that by quark pair creation mesons couple to pairs of mesons according to
processes of the form

\begin{equation}
C\;\longleftrightarrow\; A+B
\;\;\; .
\label{MMMvert}
\end{equation}

\noindent
Within multiplets of $SU_{3}$-flavour or $U_{3}$-flavour symmetry, the relative
magnitudes of the transition amplitudes for processes (~\ref{MMMvert}) are
given by

\begin{equation}
\lambda\xrm{Tr}\left({\cal M}_{A}{\cal M}_{B}{{\cal M}_{C}}^{T}\pm
{\cal M}_{B}{\cal M}_{A}{{\cal M}_{C}}^{T}\right)
\;\;\; ,
\label{simple}
\end{equation}

\noindent
where ${\cal M}_{X}$ is the $3\times 3$ flavour matrix for meson $X$.
It is understood in formula (\ref{simple}) that either the
symmetric or the antisymmetric trace is to be taken, depending on the sign of
the product of the three charge-conjugation quantum numbers. This way, charge
conjugation and $G$-parity are automatically preserved.

Relation (~\ref{simple}) is often referred to as the Zweig rule, since it
fully suppresses exactly those two-meson decay modes which do not meet
Zweig's criteria for meson-pair decay as given in Ref.~\cite{Zwe64} and
moreover agrees with the quark line rules for Quantum Chromodynamics in the
limit of large $N_{c}$ as developed in Ref.~\cite{GtH74}.

The constant $\lambda$ in front of expression (~\ref{simple}) can be
adjusted to experiment for each different set of three $U_{3}$-flavour nonets
(or octets and singlets in the case of $SU_{3}$-flavour), such that the
interaction Lagrangian for the theory may contain a long sum of all possible
three-meson vertices, each with a different coupling constant.
However, in the absence of a prescription for the relative intensities
among the thus occurring terms in the interaction Lagrangian, one easily
overlooks (see e.g.\ Ref.~\cite{Toe95}) an inconsistency in the procedure for
the scalar flavour-singlet two-meson transition modes if a unified coupling is
introduced, as we will explain in the following.
Note that, especially in the case of the scalar mesons, the employment of truly
flavour-independent couplings may be crucial for the obtainment of reliable
predictions in unitarised meson models, due to the large and highly nonlinear
coupled-channel effects in these systems \cite{Bev98a} (see also \cite{Sch98}).

\section{All three-meson vertices and specific examples}

In Ref.~\cite{Bev98b} we describe how the various values for the coupling
constants $\lambda$ can be unified. Our strategy becomes manageable, if we
assume equal effective quark masses in the harmonic-oscillator
expansion, resulting in a finite, albeit large, number of possible meson-pair
channels for each type of meson, which are all just given by the recoupling of
the four involved valence (anti)quarks. As described in Ref.~\cite{Bev98b},
the coupling constants boil down to the following expression

\begin{eqnarray}
 & & g\;\xrm{Tr}\left\{
{\cal M}_{A}{\cal M}_{B}{{\cal M}_{C}}^{T}\;
\bVk{J,L,S,N;(j,\ell ,s,n;A),(j,\ell ,s,n;B)}{\cal P}{(J,\ell ,s,n;C)}\; +
\right.\label{decint} \\ [.3cm] & & \;\;\;\;\;\;\;\;\;\; +\;\left.
{\cal M}_{B}{\cal M}_{A}{{\cal M}_{C}}^{T}\;
\bVk{J,L,S,N;(j,\ell ,s,n;A),(j,\ell ,s,n;B)}{\bar{\cal P}}{(J,\ell ,s,n;C)}
\right\},
\nonumber
\end{eqnarray}

\noindent
where $g$ is universal, i.e., the same for all possible three-meson
vertices. The quantum numbers $J$, $L$, $S$ and $N$ in formula (~\ref{decint})
represent the total angular momentum, the relative orbital angular
momentum, the total spin, and the relative radial excitation of the $A+B$
two-meson channel, respectively, whereas $(j,\ell ,s,n;X)$ represent
the corresponding quantum numbers for the $q\bar{q}$-system that describes
meson $X$. $\cal P$ represents the exchange operator for quarks and
$\bar{\cal P}$ for antiquarks. The matrix elements

\begin{eqnarray}
 & \bVk{J,L,S,N;(j,\ell ,s,n;A),(j,\ell ,s,n;B)}{\cal P}{(J,\ell ,s,n;C)} &
\xrm{(a)}\nonumber\\ [.3cm] \xrm{and}\;\;\;\; &
\bVk{J,L,S,N;(j,\ell ,s,n;B),(j,\ell ,s,n;A)}{\bar{\cal P}}{(J,\ell ,s,n;C)}
\;\;\; . & \xrm{(b)}
\label{recouple}
\end{eqnarray}

\noindent
determine the relative coupling constants for the various OZI-allowed
three-meson vertices. They result from Fermi statistics applied to the valence
(anti)quarks and Bose statistics to the meson pair in the four-particle
harmonic oscillator expansion, which is not be confused with either quark
dynamics or quark wave functions. Expressions (~\ref{recouple}a) and
(~\ref{recouple}b) are equal, up to a factor $\pm 1$, depending on the sign
of the product of the three charge-conjugation quantum numbers, which is
equivalent to the choice of sign in formula (~\ref{simple}).
Total spin $J$, parity and charge conjugation are conserved and the OZI-rule
(~\cite{Zwe64} and \cite{OI}) is respected by formula (~\ref{decint}).
The recoupling scheme is outlined in Ref.~\cite{Bev98b}, whereas more details
on the evaluation of the recoupling matrix elements (~\ref{recouple})
can be found in Ref.~\cite{Bev83a}.

In order to make our point, instead of exhibiting all details of the
calculation, we just give the results for three cases: the two-meson
transitions of pseudoscalar, vector, and scalar mesons. Table \ref{partid}
shows the nomenclature we used for the relevant mesons in this paper.
The squares of the transition amplitudes to all channels which couple
within our procedure are given in Table \ref{pseudoscalar} for pseudoscalar
mesons, in Table \ref{vector} for vector mesons, and in Table \ref{scalar} for 
scalar mesons. In order to keep the tables as condensed as possible and
since we assume that isospin is indeed a perfect symmetry, we may
represent all members of an isomultiplet by the same symbol
($t$ for isotriplet, $d$ for isodoublet, $8$ for the isosinglet
flavour-octet member, and $1$ for the flavour singlet)

Now, let us just analyse one horizontal line of one of the three tables,
to make sure that the reader understands what the numbers stand for. Let us
take the fourth line of Table \ref{pseudoscalar}. In the first column, under
$A$, we find $P$ for meson $A$, which hence characterises a meson out of the
lowest-lying pseudoscalar nonet. In the second column, under $B$, we similarly
find that meson $B$ represents a meson out of the lowest vector nonet.
In the third column, we find the quantum numbers for the relative motion of
$A$ and $B$, i.e., $P$-wave ($L=1$) with total spin one ($S=1$), in the
lowest radial excitation ($N=0$). Since the table refers to two-meson
transitions of the lowest pseudoscalar meson nonet ($P$, indicated in the top
of the table), the next four columns refer to its isotriplet member, which
is the pion. We then find that the pion couples with a strength $\sqrt{1/6}$
to the $tt$ (isotriplet-isotriplet) channel, which, following the
above-discussed particle assignments for $A$ and $B$, i.e., pseudoscalar
and vector respectively, represents the $\pi\rho$ channel.
Following a similar reasoning, we find that the pion couples
with a strength $\sqrt{1/12}$ to $KK^{\ast}$.
The total coupling of a pion to pseudoscalar-vector channels is given in the
column under $T$ by $\sqrt{1/4}$, which is the square root of the quadratic
sum of the two previous couplings, i.e., $\sqrt{1/6+1/12}$.

The next set of coupling constants refer to the two-meson transitions of a
kaon. We find $\sqrt{1/8}$ to $td$, which represents both
of the possibilities pseudoscalar (isotriplet) + vector (isodoublet),
i.e., $\pi K^{\ast}$, and pseudoscalar (isodoublet) + vector
(isotriplet), i.e., $K\rho$, each with half the intensity given in the table.
Next, we find in the table that the kaon couples with $\sqrt{1/8}$ to $d8$,
which represents both of the possibilities pseudoscalar (isodoublet) + vector
($SU_{3}$-octet isoscalar), i.e., $K$ + octet-mixture of $\omega$
and $\phi$, and pseudoscalar ($SU_{3}$-octet isoscalar) + vector
(isodoublet), i.e., octet-mixture of $\eta$ and $\eta '$ + $K^{\ast}$, each
with half the intensity given in the table.
The kaon does not couple to the $d1$ channels in pseudoscalar + vector,
which represent the channels with one isodoublet and one $SU_{3}$ singlet.
The total coupling for the kaon to its pseudoscalar + vector channels
sums up to $\sqrt{1/4}$, as one verifies in the column under $T$.
The next two sets of coupling constants similarly refer to the two-meson
transition modes of the isoscalar, either $SU_{3}$-octet or $SU_{3}$-singlet,
partners of the pseudoscalar nonet.

A remark is here in place: When we identify the flavour isotriplet members of
the lowest lying harmonic oscillator state with pions, the isodoublet states
with kaons and so on, then we have actually in mind that the corresponding
coupling constants of their three-meson vertices are to be folded in a
coupled channel model where the real mesons come out as bound states and
resonances (see e.g.\ \cite{Toe95}, \cite{Eic76} and \cite{Bev80}).
Hence, the above particle identification should not be taken
too literally. At best, one may identify the harmonic oscillator states with
objects which do not really exist in Nature, the so-called bare hadrons,
{\it i.e.} valence $q\bar{q}$-systems which are forbidden to couple to
two-meson channels by means of valence quark pair creation. Nevertheless,
the unification of the coupling constants can only be achieved by taking into
account the internal structure of the three mesons involved, for which here
we have chosen harmonic oscillators.

\section{A closer look at the results}

From the three tables we may notice the following:
\vspace{0.3cm}

{\bf 1}. The intensities (couplings squared) for all strong two-meson
transition modes of the pion (columns 4 to 7 in Table \ref{pseudoscalar})
add up to 1 (number at the very bottom of the eighth column); and the same
result holds for the couplings to the two-meson channels of all other
pseudoscalar nonet members, as well as for vector and scalar mesons
(Tables \ref{vector} and \ref{scalar}). The reason for this
property is the wave-function normalisation for the recoupling matrix elements
of formula (~\ref{decint}), which this way translates the flavour independence
of strong interactions, very recently reconfirmed by experiment \cite{Abe99}.

{\bf 2}. The subtotals (columns under $T$) for the strong two-meson transition
intensities of the octet members are equal for each different mode
(one horizontal line in each of the tables). This translates $SU_{3}$-flavour
independence of the strong interactions.

{\bf 3}. The subtotals of the flavour-singlet pseudoscalar and vector mesons
are either twice as large as those of the flavour-octet members, or zero,
in such a way that in both cases the total intensity adds up to 1.
Unfortunately, the tables for axial vectors, tensors, etc.\ are too long
to be shown here in a manageable form. Nevertheless, let us just mention
that for all higher quantum numbers we find similar factors two and zero
for the flavour-singlet couplings, with only one exception: the scalar mesons
(Table \ref{scalar}).

{\bf 4}. If one uses formula (~\ref{simple}), one similarly obtains
these factors two and zero. However, in this case the same applies to
scalar-meson transitions, \em contrary \em \/ to our findings.
\vspace{0.3cm}

In our procedure (formula (~\ref{decint})), we thus find full flavour
independence for all strong two-meson transitions of all
mesons, whereas with formula (~\ref{simple}) the flavour-singlet scalar
couples twice as strong to its two-meson channels.

This feature of the three-meson couplings can only be fully appreciated once
\em all \em \/ two-meson channels (open and closed) are taken into account,
which is much easier when their number is finite, and which takes a
particularly manageable form in the harmonic-oscillator expansion for equal
effective quark masses.

\section{Conclusions}

The flavour-singlet of scalar mesons has the quantum numbers of the vacuum
($\ket{0}$), which we also believe to be the quantum numbers of the valence
$q\bar{q}$-pair created in OZI-allowed strong two-meson transitions.
Now, in general, the normalised sum of two orthonormal states $\ket{\phi}$
and $\ket{0}$ is given by $\left(\ket{\phi}+\ket{0}\right) /\sqrt{2}$.
However, when $\ket{\phi}=\ket{0}$ (in which case they are not orthogonal),
then the correctly normalised sum is given by
$\left(\ket{\phi}+\ket{0}\right) /2$.
This is exactly the reason for the extra factor $1/\sqrt{2}$ which we find
with our procedure. Hence, we propose to modify formula (~\ref{simple}) into

\begin{equation}
\lambda\;\fnd{\xrm{Tr}\left({\cal M}_{A}{\cal M}_{B}{{\cal M}_{C}}^{T}\pm
{\cal M}_{B}{\cal M}_{A}{{\cal M}_{C}}^{T}\right)}
{\sqrt{1+\bk{C}{\xrm{flavour-singlet scalar meson}}}}
\;\;\; ,
\label{better}
\end{equation}

\noindent
in order to restore universal flavour independence for the three-meson vertices
of effective theories for strong interactions.

\section{Epilogue}

Finally, we should mention that the Zweig rule for strong decays does not
necessarily imply that singlets couple twice as strongly as octet members.
On the contrary, the couplings for three-meson vertices for both subsets of
the flavour nonet may be chosen independently in $SU_{3}$-flavour-symmetric
theories.

Moreover, we do neither assume here that the numbers of the three tables
(\ref{pseudoscalar}, \ref{vector} and \ref{scalar}) are the rigorously
correct relative intensities for three-meson vertices, nor that the number
of two-meson channels must be finite. For that, both the limit of
equal effective valence quark masses and the harmonic oscillator expansion
are probably too crude approximations. Those numbers are principally meant to
pinpoint the scalar flavour-singlet problem. Nevertheless, in the light of
the promising results of Ref.~\cite{Bev80},
in which works a unified coupling for pseudoscalars, vectors and scalars
has been applied, the tables gain some credibility. Furthermore, the
anti-De Sitter geometry \cite{Bev84c}, which has recently revived
\cite{Mal98} as a possible candidate for quark confinement \cite{Vol98a},
is well approximated by a harmonic oscillator of universal frequency. This
might provide an additional justification for the here employed
harmonic-oscillator expansion, as not just a purely mathematical tool.
\vspace{1cm}

{\it Acknowledgements.} We wish to thank H. Walliser and A. Blin for
useful discussions and suggestions.

\clearpage

\begin{table}
\begin{center}
\begin{tabular}{||c|c|c||}
\hline\hline & & \\
symbol & $(n+1)\;{^{2s+1}\ell_{J}}$ & $J^{PC}$ \\ & & \\
\hline\hline & & \\
$P$        & $1\;{^{1}S_{0}}$ & $0^{-+}$ \\ [.3cm]
$P'$       & $2\;{^{1}S_{0}}$ & $0^{-+}$ \\ [.3cm]
$V_{0}$    & $1\;{^{3}S_{1}}$ & $1^{--}$ \\ [.3cm]
${V_{0}}'$ & $2\;{^{3}S_{1}}$ & $1^{--}$ \\ [.3cm]
$V_{2}$    & $1\;{^{3}D_{1}}$ & $1^{--}$ \\ [.3cm]
$S$        & $1\;{^{3}P_{0}}$ & $0^{++}$ \\ [.3cm]
$T$        & $1\;{^{3}P_{1}}$ & $1^{++}$ \\ [.3cm]
$U$        & $1\;{^{1}P_{1}}$ & $1^{+-}$
\\ \hline\hline
\end{tabular}
\end{center}
\caption[]{Nomenclature of mesonic $q\bar{q}$ systems relevant to this paper.
The columns respectively contain our notation for the mesons, the $q\bar{q}$
quantum numbers ($n$ is radial quantum number, $s$ is total spin, $\ell$ is
orbital and $J$ is total angular momentum), and the more common quantum
numbers ($J$, parity $P=(-1)^{\ell +1}$, and charge conjugation $C=(-1)^{\ell
+s}$).}
\label{partid}
\end{table}

\begin{table}
\begin{center}
\scriptsize
\begin{tabular}{||cc|c||cccc|c||ccc|c||cccc|c||cccc|c||}
\cline{4-22} \multicolumn{3}{c||}{} & \multicolumn{19}{c||}{} \\
\multicolumn{3}{c||}{} &
\multicolumn{19}{c||}{flavour channels and totals for $P$}
 \\
\cline{4-22} \multicolumn{3}{c||}{} &
\multicolumn{14}{c||}{} & \multicolumn{5}{c||}{} \\
\multicolumn{3}{c||}{} &
\multicolumn{14}{c||}{$SU_{3}$-octet members} &
\multicolumn{5}{c||}{$SU_{3}$ singlets} \\
\hline \multicolumn{3}{||c||}{} &
\multicolumn{14}{c||}{} & \multicolumn{5}{c||}{} \\
\multicolumn{3}{||c||}{decay products} & \multicolumn{5}{c||}{isotriplets} &
\multicolumn{4}{c||}{isodoublets} &
\multicolumn{5}{c||}{isoscalars} & \multicolumn{5}{c||}{} \\ [.3cm]
$A$ & $B$ & rel. & \multicolumn{5}{c||}{($t$)} &
\multicolumn{4}{c||}{($d$)} & \multicolumn{5}{c||}{($8$)} &
\multicolumn{5}{c||}{($1$)}\\ [.3cm]
 & & $\!\! LSN\!\!$ &
$\!\! tt\!\!$ & $\!\! dd\!\!$ & $\!\! t8\!\!$ &
$\!\! t1\!\!$ & $\!\! T\!\!$ & $\!\! td\!\!$ & $\!\! d8\!\!$ &
$\!\! d1\!\!$ & $\!\! T\!\!$ & $\!\! tt\!\!$ & $\!\! dd\!\!$ &
$\!\! 88\!\!$ & $\!\! 11\!\!$ & $\!\! T\!\!$ & $\!\! tt\!\!$ &
$\!\! dd\!\!$ & $\!\! 88\!\!$ & $\!\! 11\!\!$ & $\!\! T\!\!$ \\
\hline & & & & & & & & & & & & & & & & & & & & & \\ [.3cm]
$\!\! P\!\!$ & $\!\! S\!\!$ & $\!\! 000\!\!$ &
- &
$\!\!\frac{    1}{   24}\!\!$ &
$\!\!\frac{    1}{   36}\!\!$ &
$\!\!\frac{    1}{   18}\!\!$ &
$\!\!\frac{    1}{    8}\!\!$ &
$\!\!\frac{    1}{   16}\!\!$ &
$\!\!\frac{    1}{  144}\!\!$ &
$\!\!\frac{    1}{   18}\!\!$ &
$\!\!\frac{    1}{    8}\!\!$ &
$\!\!\frac{    1}{   24}\!\!$ &
$\!\!\frac{    1}{   72}\!\!$ &
$\!\!\frac{    1}{   72}\!\!$ &
$\!\!\frac{    1}{   18}\!\!$ &
$\!\!\frac{    1}{    8}\!\!$ &
$\!\!\frac{    1}{   12}\!\!$ &
$\!\!\frac{    1}{    9}\!\!$ &
$\!\!\frac{    1}{   36}\!\!$ &
$\!\!\frac{    1}{   36}\!\!$ &
$\!\!\frac{    1}{    4}\!\!$\\ [.3cm]
$\!\! V_{0}\!\!$ & $\!\! U\!\!$ & $\!\! 000\!\!$ &
- &
$\!\!\frac{    1}{   24}\!\!$ &
$\!\!\frac{    1}{   36}\!\!$ &
$\!\!\frac{    1}{   18}\!\!$ &
$\!\!\frac{    1}{    8}\!\!$ &
$\!\!\frac{    1}{   16}\!\!$ &
$\!\!\frac{    1}{  144}\!\!$ &
$\!\!\frac{    1}{   18}\!\!$ &
$\!\!\frac{    1}{    8}\!\!$ &
$\!\!\frac{    1}{   24}\!\!$ &
$\!\!\frac{    1}{   72}\!\!$ &
$\!\!\frac{    1}{   72}\!\!$ &
$\!\!\frac{    1}{   18}\!\!$ &
$\!\!\frac{    1}{    8}\!\!$ &
$\!\!\frac{    1}{   12}\!\!$ &
$\!\!\frac{    1}{    9}\!\!$ &
$\!\!\frac{    1}{   36}\!\!$ &
$\!\!\frac{    1}{   36}\!\!$ &
$\!\!\frac{    1}{    4}\!\!$\\ [.3cm]
$\!\! V_{0}\!\!$ & $\!\! T\!\!$ & $\!\! 000\!\!$ &
$\!\!\frac{    1}{    6}\!\!$ &
$\!\!\frac{    1}{   12}\!\!$ &
- &
- &
$\!\!\frac{    1}{    4}\!\!$ &
$\!\!\frac{    1}{    8}\!\!$ &
$\!\!\frac{    1}{    8}\!\!$ &
- &
$\!\!\frac{    1}{    4}\!\!$ &
- &
$\!\!\frac{    1}{    4}\!\!$ &
- &
- &
$\!\!\frac{    1}{    4}\!\!$ &
- &
- &
- &
- &
- \\ [.3cm]
$\!\! P\!\!$ & $\!\! V_{0}\!\!$ & $\!\! 110\!\!$ &
$\!\!\frac{    1}{    6}\!\!$ &
$\!\!\frac{    1}{   12}\!\!$ &
- &
- &
$\!\!\frac{    1}{    4}\!\!$ &
$\!\!\frac{    1}{    8}\!\!$ &
$\!\!\frac{    1}{    8}\!\!$ &
- &
$\!\!\frac{    1}{    4}\!\!$ &
- &
$\!\!\frac{    1}{    4}\!\!$ &
- &
- &
$\!\!\frac{    1}{    4}\!\!$ &
- &
- &
- &
- &
- \\ [.3cm]
$\!\! V_{0}\!\!$ & $\!\! V_{0}\!\!$ & $\!\! 110\!\!$ &
- &
$\!\!\frac{    1}{   12}\!\!$ &
$\!\!\frac{    1}{   18}\!\!$ &
$\!\!\frac{    1}{    9}\!\!$ &
$\!\!\frac{    1}{    4}\!\!$ &
$\!\!\frac{    1}{    8}\!\!$ &
$\!\!\frac{    1}{   72}\!\!$ &
$\!\!\frac{    1}{    9}\!\!$ &
$\!\!\frac{    1}{    4}\!\!$ &
$\!\!\frac{    1}{   12}\!\!$ &
$\!\!\frac{    1}{   36}\!\!$ &
$\!\!\frac{    1}{   36}\!\!$ &
$\!\!\frac{    1}{    9}\!\!$ &
$\!\!\frac{    1}{    4}\!\!$ &
$\!\!\frac{    1}{    6}\!\!$ &
$\!\!\frac{    2}{    9}\!\!$ &
$\!\!\frac{    1}{   18}\!\!$ &
$\!\!\frac{    1}{   18}\!\!$ &
$\!\!\frac{    1}{    2}\!\!$
 \\ & & & & & & & & & & & & & & & & & & & & & \\ \hline
\multicolumn{7}{c|}{} & & \multicolumn{3}{c|}{} & &
\multicolumn{4}{c|}{} & & \multicolumn{4}{c|}{} & \\
\multicolumn{7}{c|}{} & 1 & \multicolumn{3}{c|}{} & 1 &
\multicolumn{4}{c|}{} & 1 & \multicolumn{4}{c|}{} & 1 \\
\cline{8-8} \cline{12-12} \cline{17-17} \cline{22-22}
\end{tabular}
\normalsize
\end{center}
\caption[]{Transition intensities for the coupling of pseudoscalar
mesons to meson pairs. The interpretation of the content of the table is
explained in the text.}
\label{pseudoscalar}
\end{table}

\begin{table}
\begin{center}
\scriptsize
\begin{tabular}{||cc|c||cccc|c||ccc|c||cccc|c||cccc|c||}
\cline{4-22} \multicolumn{3}{c||}{} & \multicolumn{19}{c||}{} \\
\multicolumn{3}{c||}{} &
\multicolumn{19}{c||}{flavour channels and totals for $V_{0}$}\\
\cline{4-22} \multicolumn{3}{c||}{} &
\multicolumn{14}{c||}{} & \multicolumn{5}{c||}{} \\
\multicolumn{3}{c||}{} &
\multicolumn{14}{c||}{$SU_{3}$-octet members} &
\multicolumn{5}{c||}{$SU_{3}$ singlets} \\
\hline \multicolumn{3}{||c||}{} &
\multicolumn{14}{c||}{} & \multicolumn{5}{c||}{} \\
\multicolumn{3}{||c||}{decay products} & \multicolumn{5}{c||}{isotriplets} &
\multicolumn{4}{c||}{isodoublets} &
\multicolumn{5}{c||}{isoscalars} & \multicolumn{5}{c||}{} \\ [.3cm]
$A$ & $B$ & rel. & \multicolumn{5}{c||}{($t$)} &
\multicolumn{4}{c||}{($d$)} & \multicolumn{5}{c||}{($8$)} &
\multicolumn{5}{c||}{($1$)}\\ [.3cm]
 & & $\!\! LSN\!\!$ &
$\!\! tt\!\!$ & $\!\! dd\!\!$ & $\!\! t8\!\!$ &
$\!\! t1\!\!$ & $\!\! T\!\!$ & $\!\! td\!\!$ & $\!\! d8\!\!$ &
$\!\! d1\!\!$ & $\!\! T\!\!$ & $\!\! tt\!\!$ & $\!\! dd\!\!$ &
$\!\! 88\!\!$ & $\!\! 11\!\!$ & $\!\! T\!\!$ & $\!\! tt\!\!$ &
$\!\! dd\!\!$ & $\!\! 88\!\!$ & $\!\! 11\!\!$ & $\!\! T\!\!$ \\
\hline & & & & & & & & & & & & & & & & & & & & & \\ [.3cm]
$\!\! P\!\!$ & $\!\! U\!\!$ & $\!\! 010\!\!$ &
- &
$\!\!\frac{    1}{   72}\!\!$ &
$\!\!\frac{    1}{  108}\!\!$ &
$\!\!\frac{    1}{   54}\!\!$ &
$\!\!\frac{    1}{   24}\!\!$ &
$\!\!\frac{    1}{   48}\!\!$ &
$\!\!\frac{    1}{  432}\!\!$ &
$\!\!\frac{    1}{   54}\!\!$ &
$\!\!\frac{    1}{   24}\!\!$ &
$\!\!\frac{    1}{   72}\!\!$ &
$\!\!\frac{    1}{  216}\!\!$ &
$\!\!\frac{    1}{  216}\!\!$ &
$\!\!\frac{    1}{   54}\!\!$ &
$\!\!\frac{    1}{   24}\!\!$ &
$\!\!\frac{    1}{   36}\!\!$ &
$\!\!\frac{    1}{   27}\!\!$ &
$\!\!\frac{    1}{  108}\!\!$ &
$\!\!\frac{    1}{  108}\!\!$ &
$\!\!\frac{    1}{   12}\!\!$\\ [.3cm]
$\!\! P\!\!$ & $\!\! T\!\!$ & $\!\! 010\!\!$ &
$\!\!\frac{    1}{   18}\!\!$ &
$\!\!\frac{    1}{   36}\!\!$ &
- &
- &
$\!\!\frac{    1}{   12}\!\!$ &
$\!\!\frac{    1}{   24}\!\!$ &
$\!\!\frac{    1}{   24}\!\!$ &
- &
$\!\!\frac{    1}{   12}\!\!$ &
- &
$\!\!\frac{    1}{   12}\!\!$ &
- &
- &
$\!\!\frac{    1}{   12}\!\!$ &
- &
- &
- &
- &
- \\ [.3cm]
$\!\! V_{0}\!\!$ & $\!\! U\!\!$ & $\!\! 010\!\!$ &
$\!\!\frac{    1}{   18}\!\!$ &
$\!\!\frac{    1}{   36}\!\!$ &
- &
- &
$\!\!\frac{    1}{   12}\!\!$ &
$\!\!\frac{    1}{   24}\!\!$ &
$\!\!\frac{    1}{   24}\!\!$ &
- &
$\!\!\frac{    1}{   12}\!\!$ &
- &
$\!\!\frac{    1}{   12}\!\!$ &
- &
- &
$\!\!\frac{    1}{   12}\!\!$ &
- &
- &
- &
- &
- \\ [.3cm]
$\!\! V_{0}\!\!$ & $\!\! T\!\!$ & $\!\! 010\!\!$ &
- &
$\!\!\frac{    1}{   18}\!\!$ &
$\!\!\frac{    1}{   27}\!\!$ &
$\!\!\frac{    2}{   27}\!\!$ &
$\!\!\frac{    1}{    6}\!\!$ &
$\!\!\frac{    1}{   12}\!\!$ &
$\!\!\frac{    1}{  108}\!\!$ &
$\!\!\frac{    2}{   27}\!\!$ &
$\!\!\frac{    1}{    6}\!\!$ &
$\!\!\frac{    1}{   18}\!\!$ &
$\!\!\frac{    1}{   54}\!\!$ &
$\!\!\frac{    1}{   54}\!\!$ &
$\!\!\frac{    2}{   27}\!\!$ &
$\!\!\frac{    1}{    6}\!\!$ &
$\!\!\frac{    1}{    9}\!\!$ &
$\!\!\frac{    4}{   27}\!\!$ &
$\!\!\frac{    1}{   27}\!\!$ &
$\!\!\frac{    1}{   27}\!\!$ &
$\!\!\frac{    1}{    3}\!\!$\\ [.3cm]
$\!\! S\!\!$ & $\!\! V_{0}\!\!$ & $\!\! 010\!\!$ &
- &
$\!\!\frac{    1}{   24}\!\!$ &
$\!\!\frac{    1}{   36}\!\!$ &
$\!\!\frac{    1}{   18}\!\!$ &
$\!\!\frac{    1}{    8}\!\!$ &
$\!\!\frac{    1}{   16}\!\!$ &
$\!\!\frac{    1}{  144}\!\!$ &
$\!\!\frac{    1}{   18}\!\!$ &
$\!\!\frac{    1}{    8}\!\!$ &
$\!\!\frac{    1}{   24}\!\!$ &
$\!\!\frac{    1}{   72}\!\!$ &
$\!\!\frac{    1}{   72}\!\!$ &
$\!\!\frac{    1}{   18}\!\!$ &
$\!\!\frac{    1}{    8}\!\!$ &
$\!\!\frac{    1}{   12}\!\!$ &
$\!\!\frac{    1}{    9}\!\!$ &
$\!\!\frac{    1}{   36}\!\!$ &
$\!\!\frac{    1}{   36}\!\!$ &
$\!\!\frac{    1}{    4}\!\!$\\ [.3cm]
$\!\! P\!\!$ & $\!\! P\!\!$ & $\!\! 100\!\!$ &
$\!\!\frac{    1}{   36}\!\!$ &
$\!\!\frac{    1}{   72}\!\!$ &
- &
- &
$\!\!\frac{    1}{   24}\!\!$ &
$\!\!\frac{    1}{   48}\!\!$ &
$\!\!\frac{    1}{   48}\!\!$ &
- &
$\!\!\frac{    1}{   24}\!\!$ &
- &
$\!\!\frac{    1}{   24}\!\!$ &
- &
- &
$\!\!\frac{    1}{   24}\!\!$ &
- &
- &
- &
- &
- \\ [.3cm]
$\!\! V_{0}\!\!$ & $\!\! V_{0}\!\!$ & $\!\! 100\!\!$ &
$\!\!\frac{    1}{  108}\!\!$ &
$\!\!\frac{    1}{  216}\!\!$ &
- &
- &
$\!\!\frac{    1}{   72}\!\!$ &
$\!\!\frac{    1}{  144}\!\!$ &
$\!\!\frac{    1}{  144}\!\!$ &
- &
$\!\!\frac{    1}{   72}\!\!$ &
- &
$\!\!\frac{    1}{   72}\!\!$ &
- &
- &
$\!\!\frac{    1}{   72}\!\!$ &
- &
- &
- &
- &
- \\ [.3cm]
$\!\! P\!\!$ & $\!\! V_{0}\!\!$ & $\!\! 110\!\!$ &
- &
$\!\!\frac{    1}{   18}\!\!$ &
$\!\!\frac{    1}{   27}\!\!$ &
$\!\!\frac{    2}{   27}\!\!$ &
$\!\!\frac{    1}{    6}\!\!$ &
$\!\!\frac{    1}{   12}\!\!$ &
$\!\!\frac{    1}{  108}\!\!$ &
$\!\!\frac{    2}{   27}\!\!$ &
$\!\!\frac{    1}{    6}\!\!$ &
$\!\!\frac{    1}{   18}\!\!$ &
$\!\!\frac{    1}{   54}\!\!$ &
$\!\!\frac{    1}{   54}\!\!$ &
$\!\!\frac{    2}{   27}\!\!$ &
$\!\!\frac{    1}{    6}\!\!$ &
$\!\!\frac{    1}{    9}\!\!$ &
$\!\!\frac{    4}{   27}\!\!$ &
$\!\!\frac{    1}{   27}\!\!$ &
$\!\!\frac{    1}{   27}\!\!$ &
$\!\!\frac{    1}{    3}\!\!$\\ [.3cm]
$\!\! V_{0}\!\!$ & $\!\! V_{0}\!\!$ & $\!\! 120\!\!$ &
$\!\!\frac{    5}{   27}\!\!$ &
$\!\!\frac{    5}{   54}\!\!$ &
- &
- &
$\!\!\frac{    5}{   18}\!\!$ &
$\!\!\frac{    5}{   36}\!\!$ &
$\!\!\frac{    5}{   36}\!\!$ &
- &
$\!\!\frac{    5}{   18}\!\!$ &
- &
$\!\!\frac{    5}{   18}\!\!$ &
- &
- &
$\!\!\frac{    5}{   18}\!\!$ &
- &
- &
- &
- &
- \\ & & & & & & & & & & & & & & & & & & & & & \\ \hline
\multicolumn{7}{c|}{} & & \multicolumn{3}{c|}{} & &
\multicolumn{4}{c|}{} & & \multicolumn{4}{c|}{} & \\
\multicolumn{7}{c|}{} & 1 & \multicolumn{3}{c|}{} & 1 &
\multicolumn{4}{c|}{} & 1 & \multicolumn{4}{c|}{} & 1 \\
\cline{8-8} \cline{12-12} \cline{17-17} \cline{22-22}
\end{tabular}
\normalsize
\end{center}
\caption[]{Transition intensities for the coupling of vector
mesons to meson pairs.}
\label{vector}
\end{table}

\begin{table}
\begin{center}
\scriptsize
\begin{tabular}{||cc|c||cccc|c||ccc|c||cccc|c||cccc|c||}
\cline{4-22} \multicolumn{3}{c||}{} & \multicolumn{19}{c||}{} \\
\multicolumn{3}{c||}{} &
\multicolumn{19}{c||}{flavour channels and totals for $S$}\\
\cline{4-22} \multicolumn{3}{c||}{} &
\multicolumn{14}{c||}{} & \multicolumn{5}{c||}{} \\
\multicolumn{3}{c||}{} &
\multicolumn{14}{c||}{$SU_{3}$-octet members} &
\multicolumn{5}{c||}{$SU_{3}$ singlets} \\
\hline \multicolumn{3}{||c||}{} &
\multicolumn{14}{c||}{} & \multicolumn{5}{c||}{} \\
\multicolumn{3}{||c||}{decay products} & \multicolumn{5}{c||}{isotriplets} &
\multicolumn{4}{c||}{isodoublets} &
\multicolumn{5}{c||}{isoscalars} & \multicolumn{5}{c||}{} \\ [.3cm]
$A$ & $B$ & rel. & \multicolumn{5}{c||}{($t$)} &
\multicolumn{4}{c||}{($d$)} & \multicolumn{5}{c||}{($8$)} &
\multicolumn{5}{c||}{($1$)}\\ [.3cm]
 & & $\!\! LSN\!\!$ &
$\!\! tt\!\!$ & $\!\! dd\!\!$ & $\!\! t8\!\!$ &
$\!\! t1\!\!$ & $\!\! T\!\!$ & $\!\! td\!\!$ & $\!\! d8\!\!$ &
$\!\! d1\!\!$ & $\!\! T\!\!$ & $\!\! tt\!\!$ & $\!\! dd\!\!$ &
$\!\! 88\!\!$ & $\!\! 11\!\!$ & $\!\! T\!\!$ & $\!\! tt\!\!$ &
$\!\! dd\!\!$ & $\!\! 88\!\!$ & $\!\! 11\!\!$ & $\!\! T\!\!$ \\
\hline & & & & & & & & & & & & & & & & & & & & & \\ [.3cm]
$\!\! P\!\!$ & $\!\! P\!\!$ & $\!\! 001\!\!$ &
- &
$\!\!\frac{    1}{   72}\!\!$ &
$\!\!\frac{    1}{  108}\!\!$ &
$\!\!\frac{    1}{   54}\!\!$ &
$\!\!\frac{    1}{   24}\!\!$ &
$\!\!\frac{    1}{   48}\!\!$ &
$\!\!\frac{    1}{  432}\!\!$ &
$\!\!\frac{    1}{   54}\!\!$ &
$\!\!\frac{    1}{   24}\!\!$ &
$\!\!\frac{    1}{   72}\!\!$ &
$\!\!\frac{    1}{  216}\!\!$ &
$\!\!\frac{    1}{  216}\!\!$ &
$\!\!\frac{    1}{   54}\!\!$ &
$\!\!\frac{    1}{   24}\!\!$ &
$\!\!\frac{    1}{   72}\!\!$ &
$\!\!\frac{    1}{   54}\!\!$ &
$\!\!\frac{    1}{  216}\!\!$ &
$\!\!\frac{    1}{  216}\!\!$ &
$\!\!\frac{    1}{   24}\!\!$\\ [.3cm]
$\!\! P\!\!$ & $\!\! P'\!\!$ & $\!\! 000\!\!$ &
- &
$\!\!\frac{    1}{  144}\!\!$ &
$\!\!\frac{    1}{  216}\!\!$ &
$\!\!\frac{    1}{  108}\!\!$ &
$\!\!\frac{    1}{   48}\!\!$ &
$\!\!\frac{    1}{   96}\!\!$ &
$\!\!\frac{    1}{  864}\!\!$ &
$\!\!\frac{    1}{  108}\!\!$ &
$\!\!\frac{    1}{   48}\!\!$ &
$\!\!\frac{    1}{  144}\!\!$ &
$\!\!\frac{    1}{  432}\!\!$ &
$\!\!\frac{    1}{  432}\!\!$ &
$\!\!\frac{    1}{  108}\!\!$ &
$\!\!\frac{    1}{   48}\!\!$ &
$\!\!\frac{    1}{  144}\!\!$ &
$\!\!\frac{    1}{  108}\!\!$ &
$\!\!\frac{    1}{  432}\!\!$ &
$\!\!\frac{    1}{  432}\!\!$ &
$\!\!\frac{    1}{   48}\!\!$\\ [.3cm]
$\!\! V_{0}\!\!$ & $\!\! V_{0}\!\!$ & $\!\! 001\!\!$ &
- &
$\!\!\frac{    1}{  216}\!\!$ &
$\!\!\frac{    1}{  324}\!\!$ &
$\!\!\frac{    1}{  162}\!\!$ &
$\!\!\frac{    1}{   72}\!\!$ &
$\!\!\frac{    1}{  144}\!\!$ &
$\!\!\frac{    1}{ 1296}\!\!$ &
$\!\!\frac{    1}{  162}\!\!$ &
$\!\!\frac{    1}{   72}\!\!$ &
$\!\!\frac{    1}{  216}\!\!$ &
$\!\!\frac{    1}{  648}\!\!$ &
$\!\!\frac{    1}{  648}\!\!$ &
$\!\!\frac{    1}{  162}\!\!$ &
$\!\!\frac{    1}{   72}\!\!$ &
$\!\!\frac{    1}{  216}\!\!$ &
$\!\!\frac{    1}{  162}\!\!$ &
$\!\!\frac{    1}{  648}\!\!$ &
$\!\!\frac{    1}{  648}\!\!$ &
$\!\!\frac{    1}{   72}\!\!$\\ [.3cm]
$\!\! V_{0}\!\!$ & $\!\! {V_{0}}'\!\!$ & $\!\! 000\!\!$ &
- &
$\!\!\frac{    1}{  432}\!\!$ &
$\!\!\frac{    1}{  648}\!\!$ &
$\!\!\frac{    1}{  324}\!\!$ &
$\!\!\frac{    1}{  144}\!\!$ &
$\!\!\frac{    1}{  288}\!\!$ &
$\!\!\frac{    1}{ 2592}\!\!$ &
$\!\!\frac{    1}{  324}\!\!$ &
$\!\!\frac{    1}{  144}\!\!$ &
$\!\!\frac{    1}{  432}\!\!$ &
$\!\!\frac{    1}{ 1296}\!\!$ &
$\!\!\frac{    1}{ 1296}\!\!$ &
$\!\!\frac{    1}{  324}\!\!$ &
$\!\!\frac{    1}{  144}\!\!$ &
$\!\!\frac{    1}{  432}\!\!$ &
$\!\!\frac{    1}{  324}\!\!$ &
$\!\!\frac{    1}{ 1296}\!\!$ &
$\!\!\frac{    1}{ 1296}\!\!$ &
$\!\!\frac{    1}{  144}\!\!$\\ [.3cm]
$\!\! V_{0}\!\!$ & $\!\! V_{2}\!\!$ & $\!\! 000\!\!$ &
- &
$\!\!\frac{    5}{  108}\!\!$ &
$\!\!\frac{    5}{  162}\!\!$ &
$\!\!\frac{    5}{   81}\!\!$ &
$\!\!\frac{    5}{   36}\!\!$ &
$\!\!\frac{    5}{   72}\!\!$ &
$\!\!\frac{    5}{  648}\!\!$ &
$\!\!\frac{    5}{   81}\!\!$ &
$\!\!\frac{    5}{   36}\!\!$ &
$\!\!\frac{    5}{  108}\!\!$ &
$\!\!\frac{    5}{  324}\!\!$ &
$\!\!\frac{    5}{  324}\!\!$ &
$\!\!\frac{    5}{   81}\!\!$ &
$\!\!\frac{    5}{   36}\!\!$ &
$\!\!\frac{    5}{  108}\!\!$ &
$\!\!\frac{    5}{   81}\!\!$ &
$\!\!\frac{    5}{  324}\!\!$ &
$\!\!\frac{    5}{  324}\!\!$ &
$\!\!\frac{    5}{   36}\!\!$\\ [.3cm]
$\!\! U\!\!$ & $\!\! U\!\!$ & $\!\! 000\!\!$ &
- &
$\!\!\frac{    1}{  144}\!\!$ &
$\!\!\frac{    1}{  216}\!\!$ &
$\!\!\frac{    1}{  108}\!\!$ &
$\!\!\frac{    1}{   48}\!\!$ &
$\!\!\frac{    1}{   96}\!\!$ &
$\!\!\frac{    1}{  864}\!\!$ &
$\!\!\frac{    1}{  108}\!\!$ &
$\!\!\frac{    1}{   48}\!\!$ &
$\!\!\frac{    1}{  144}\!\!$ &
$\!\!\frac{    1}{  432}\!\!$ &
$\!\!\frac{    1}{  432}\!\!$ &
$\!\!\frac{    1}{  108}\!\!$ &
$\!\!\frac{    1}{   48}\!\!$ &
$\!\!\frac{    1}{  144}\!\!$ &
$\!\!\frac{    1}{  108}\!\!$ &
$\!\!\frac{    1}{  432}\!\!$ &
$\!\!\frac{    1}{  432}\!\!$ &
$\!\!\frac{    1}{   48}\!\!$\\ [.3cm]
$\!\! S\!\!$ & $\!\! S\!\!$ & $\!\! 000\!\!$ &
- &
$\!\!\frac{    1}{   48}\!\!$ &
$\!\!\frac{    1}{   72}\!\!$ &
$\!\!\frac{    1}{   36}\!\!$ &
$\!\!\frac{    1}{   16}\!\!$ &
$\!\!\frac{    1}{   32}\!\!$ &
$\!\!\frac{    1}{  288}\!\!$ &
$\!\!\frac{    1}{   36}\!\!$ &
$\!\!\frac{    1}{   16}\!\!$ &
$\!\!\frac{    1}{   48}\!\!$ &
$\!\!\frac{    1}{  144}\!\!$ &
$\!\!\frac{    1}{  144}\!\!$ &
$\!\!\frac{    1}{   36}\!\!$ &
$\!\!\frac{    1}{   16}\!\!$ &
$\!\!\frac{    1}{   48}\!\!$ &
$\!\!\frac{    1}{   36}\!\!$ &
$\!\!\frac{    1}{  144}\!\!$ &
$\!\!\frac{    1}{  144}\!\!$ &
$\!\!\frac{    1}{   16}\!\!$\\ [.3cm]
$\!\! T\!\!$ & $\!\! T\!\!$ & $\!\! 000\!\!$ &
- &
$\!\!\frac{    1}{   36}\!\!$ &
$\!\!\frac{    1}{   54}\!\!$ &
$\!\!\frac{    1}{   27}\!\!$ &
$\!\!\frac{    1}{   12}\!\!$ &
$\!\!\frac{    1}{   24}\!\!$ &
$\!\!\frac{    1}{  216}\!\!$ &
$\!\!\frac{    1}{   27}\!\!$ &
$\!\!\frac{    1}{   12}\!\!$ &
$\!\!\frac{    1}{   36}\!\!$ &
$\!\!\frac{    1}{  108}\!\!$ &
$\!\!\frac{    1}{  108}\!\!$ &
$\!\!\frac{    1}{   27}\!\!$ &
$\!\!\frac{    1}{   12}\!\!$ &
$\!\!\frac{    1}{   36}\!\!$ &
$\!\!\frac{    1}{   27}\!\!$ &
$\!\!\frac{    1}{  108}\!\!$ &
$\!\!\frac{    1}{  108}\!\!$ &
$\!\!\frac{    1}{   12}\!\!$\\ [.3cm]
$\!\! P\!\!$ & $\!\! T\!\!$ & $\!\! 110\!\!$ &
- &
$\!\!\frac{    1}{   18}\!\!$ &
$\!\!\frac{    1}{   27}\!\!$ &
$\!\!\frac{    2}{   27}\!\!$ &
$\!\!\frac{    1}{    6}\!\!$ &
$\!\!\frac{    1}{   12}\!\!$ &
$\!\!\frac{    1}{  108}\!\!$ &
$\!\!\frac{    2}{   27}\!\!$ &
$\!\!\frac{    1}{    6}\!\!$ &
$\!\!\frac{    1}{   18}\!\!$ &
$\!\!\frac{    1}{   54}\!\!$ &
$\!\!\frac{    1}{   54}\!\!$ &
$\!\!\frac{    2}{   27}\!\!$ &
$\!\!\frac{    1}{    6}\!\!$ &
$\!\!\frac{    1}{   18}\!\!$ &
$\!\!\frac{    2}{   27}\!\!$ &
$\!\!\frac{    1}{   54}\!\!$ &
$\!\!\frac{    1}{   54}\!\!$ &
$\!\!\frac{    1}{    6}\!\!$\\ [.3cm]
$\!\! V_{0}\!\!$ & $\!\! U\!\!$ & $\!\! 110\!\!$ &
- &
$\!\!\frac{    1}{   18}\!\!$ &
$\!\!\frac{    1}{   27}\!\!$ &
$\!\!\frac{    2}{   27}\!\!$ &
$\!\!\frac{    1}{    6}\!\!$ &
$\!\!\frac{    1}{   12}\!\!$ &
$\!\!\frac{    1}{  108}\!\!$ &
$\!\!\frac{    2}{   27}\!\!$ &
$\!\!\frac{    1}{    6}\!\!$ &
$\!\!\frac{    1}{   18}\!\!$ &
$\!\!\frac{    1}{   54}\!\!$ &
$\!\!\frac{    1}{   54}\!\!$ &
$\!\!\frac{    2}{   27}\!\!$ &
$\!\!\frac{    1}{    6}\!\!$ &
$\!\!\frac{    1}{   18}\!\!$ &
$\!\!\frac{    2}{   27}\!\!$ &
$\!\!\frac{    1}{   54}\!\!$ &
$\!\!\frac{    1}{   54}\!\!$ &
$\!\!\frac{    1}{    6}\!\!$\\ [.3cm]
$\!\! V_{0}\!\!$ & $\!\! V_{0}\!\!$ & $\!\! 220\!\!$ &
- &
$\!\!\frac{    5}{   54}\!\!$ &
$\!\!\frac{    5}{   81}\!\!$ &
$\!\!\frac{   10}{   81}\!\!$ &
$\!\!\frac{    5}{   18}\!\!$ &
$\!\!\frac{    5}{   36}\!\!$ &
$\!\!\frac{    5}{  324}\!\!$ &
$\!\!\frac{   10}{   81}\!\!$ &
$\!\!\frac{    5}{   18}\!\!$ &
$\!\!\frac{    5}{   54}\!\!$ &
$\!\!\frac{    5}{  162}\!\!$ &
$\!\!\frac{    5}{  162}\!\!$ &
$\!\!\frac{   10}{   81}\!\!$ &
$\!\!\frac{    5}{   18}\!\!$ &
$\!\!\frac{    5}{   54}\!\!$ &
$\!\!\frac{   10}{   81}\!\!$ &
$\!\!\frac{    5}{  162}\!\!$ &
$\!\!\frac{    5}{  162}\!\!$ &
$\!\!\frac{    5}{   18}\!\!$
 \\ & & & & & & & & & & & & & & & & & & & & & \\ \hline
\multicolumn{7}{c|}{} & & \multicolumn{3}{c|}{} & &
\multicolumn{4}{c|}{} & & \multicolumn{4}{c|}{} & \\
\multicolumn{7}{c|}{} & 1 & \multicolumn{3}{c|}{} & 1 &
\multicolumn{4}{c|}{} & 1 & \multicolumn{4}{c|}{} & 1 \\
\cline{8-8} \cline{12-12} \cline{17-17} \cline{22-22}
\end{tabular}
\normalsize
\end{center}
\caption[]{Transition intensities for the coupling of scalar
mesons to meson pairs.}
\label{scalar}
\end{table}

\clearpage

\begin{thebibliography}{10}

\bibitem{Zwe64}                 
G.~Zweig, {\it An $SU_{3}$ model for strong interaction symmetry and its
breaking},
CERN Reports TH-401 and TH-412 (1964).

\bibitem{GtH74}                 
G.~'t~Hooft,
{Nucl.\ Phys.} {\bf B72}, 461(1974);
G.~Veneziano,
{Nucl.\ Phys.} {\bf B117}, 519 (1976);
E.~Witten,
{Nucl.\ Phys.} {\bf B156}, 269 (1979).

\bibitem{Toe95}                 
Nils~A. T\"{o}rnqvist, {Zeit.\ Phys.} {\bf C68}, 647 (1995).

\bibitem{Bev98a}                 
Eef van Beveren and George Rupp, {\it Comment on {\sl ``Understanding the
scalar meson $q\bar{q}$ nonet''}}, Preprint hep-ph/9806246 (1998), submitted
for publication.

\bibitem{Sch98}
Deirdre Black, Amir H. Fariborz, Francesco Sannino and Joseph Schechter,
{Phys.\ Rev.} {D58}, 054012 (1998); {\it Putative light scalar nonet},
Preprint hep-ph/9808415 (1998), submitted for publication in {Phys.\ Rev.}
{\bf D};
Amir H. Fariborz and Joseph Schechter, {\it $\eta'\rightarrow\eta\pi\pi$
decay as a probe of a possible lowest-lying scalar nonet},
Preprint hep-ph/9902238 (1999);
Shin Ishida, Muneyuki Ishida, Taku Ishida, Kunio Takamatsu and Tsuneaki Tsuru,
{Prog.\ Theor.\ Phys.} {\bf 95}, 745 (1996);
M. Napsuciale, {\it Scalar meson masses and mixing angle in a
$U_{3}\times U_{3}$ linear sigma model}, Preprint hep-ph/9803396 (1998).

\bibitem{Bev98b}                 
Eef van Beveren and George Rupp, {\it Flavour symmetry of mesonic decay
couplings}, Preprint hep-ph/9806248 (1998), submitted for publication.

\bibitem{OI}                     
S. Okubo, {Phys.\ Lett.} {\bf 5}, 105 (1963);
J. Iizuka, K. Okada and O. Shito, {Prog.\ Theor.\ Phys.} {\bf 35}, 1061
(1966).

\bibitem{Bev83a}                 
E.~van Beveren, {Zeit.\ Phys.} {\bf C17}, 135 (1983); {\bf C21}, 291 (1984).

\bibitem{Eic76}
E.~Eichten et~al., {Phys.\ Rev.\ Lett.} {\bf 36}, 500 (1976);
{Phys.\ Rev.} {\bf D21}, 203 (1980);
Nils~A. T\"{o}rnqvist, {Phys.\ Rev.\ Lett.} {\bf 49}, 624 (1982);
Nils~A. T\"{o}rnqvist and Matts Roos, {Phys.\ Rev.\ Lett.} {\bf 76}, 1575
(1996);
Veronique Bernard, Ulf. G. Meissner, Alex Blin and Brigitte Hiller,
{Phys.\ Lett.} {\bf B253}, 443 (1991).

\bibitem{Bev80}                 
E.~van Beveren, C.~Dullemond, and G.~Rupp,
{Phys.\ Rev.} {\bf D21}, 772; (E) {\bf D22}, 787 (1980);
E.~van Beveren, G.~Rupp, T.A. Rijken, and C.~Dullemond,
{Phys.\ Rev.} {\bf D27}, 1527 (1983);
E.~van Beveren et~al., {Zeit.\ Phys.} {\bf C30}, 615 (1986).

\bibitem{Abe99}                 
K.~Abe et~al.,
{Phys.\ Rev.} {\bf D59}, 012002 (1999).

\bibitem{Bev84c}                 
E.~van Beveren, C.~Dullemond, and T.A. Rijken,
{Phys.\ Rev.} {\bf D30}, 1103 (1984);
E.~van Beveren, T.A. Rijken, C.~Dullemond, and G.~Rupp,
{\it Geometric quark confinement and hadronic resonances},
in S.~Albeverio, L.~S.~Ferreira, and L.~Streit, editors, {\em Resonances -
Models and Phenomena}, volume Lecture Notes in Physics {\bf 211}, pages
331--346, Bielefeld (1984);
C.~Dullemond, T.A. Rijken, and E.~van Beveren,
{Il Nuovo Cim.} {\bf 80A}, 401 (1984);
E.~van Beveren, T.A. Rijken, and C.~Dullemond,
{J.\ Math.\ Phys.} {\bf 27}, 1411 (1986).

\bibitem{Mal98}                 
Juan~Mart\'{\i}n Maldacena,
{Adv.\ Theor.\ Math.\ Phys.} {\bf 2}, 231 (1998).

\bibitem{Vol98a}                 
I.~V.~Volovich, {\it Large-N gauge theories and the anti-De Sitter bag model},
Preprint hep-th/9803174 (1998); Nicholas Dorey, Timothy J. Hollowood,
Valentin V. Khoze, Michael P. Mattis and Stefan Vandoren, {\it Multi-instanton
calculus and the aDS/CFT correspondence in $N=4$ superconformal field theory},
Preprint hep-th/9901128 (1999).

\end{thebibliography}
\end{document}